\documentclass[aps,prb,twocolumn,groupedaddress,showpacs]{revtex4}
\usepackage{amssymb}
\usepackage[fleqn]{amsmath}
\usepackage[dvips]{graphicx}
\usepackage{docs}
\usepackage{bm}
\usepackage{textcomp}
\usepackage[colorlinks=true,linkcolor=blue,citecolor=blue,urlcolor=black]{hyperref}%
\expandafter\ifx\csname package@font\endcsname\relax\else
 \expandafter\expandafter
 \expandafter\usepackage
 \expandafter\expandafter
 \expandafter{\csname package@font\endcsname}%
\fi

\usepackage[papersize={9in,12in}]{geometry}
\newlength{\Figwidth}
\begin{document}
\title{Structural and Electrical Properties of Bilayer SiX (X= N, P, As and Sb)}

\author{Nayereh Ghobadi}
\affiliation{Department of Electrical Engineering, University of Zanjan, Zanjan, Iran}
\author{Shoeib Babaee Touski}
\email{touski@hut.ac.ir}
\affiliation{Department of Electrical Engineering, Hamedan University of Technology, Hamedan, Iran}

\begin{abstract}
	In this work, the structural, electrical, and optical properties of bilayer SiX (X= N, P, As, and Sb) are studied using density functional theory (DFT). Five different stacking orders are considered for every compound and their structural properties are presented. The band structure of these materials demonstrates that they are indirect semiconductors. The out-of-plane strain has been applied to tune the bandgap and its electrical properties. The bandgap increases with tensile strain, whereas, compressive strain leads to semiconductor-to-metal transition. The sensitivity of the bandgap to the pressure is investigated and bilayer SiSb demonstrates the highest bandgap sensitivity to the pressure. These structures exhibit Mexican hat-like valence band dispersion that can be approved by a singularity in the density of states. The Mexican-hat coefficient can be tuned by out-of-plane strain. Optical absorption of these compounds shows that the second and lower valence bands due to the high density of states display a higher contribution to optical transitions. 
\end{abstract}



\maketitle

\section{Introduction}

Two dimensional (2D) materials have become the head of research after exfoliation of graphene\cite{novoselov2004electric}. The 2D structures of the other members of the group-IV atoms such as silicene, germanene, stanene, and plumbene have been reported theoretically and experimentally \cite{cahangirov2009two,xu2013large,yu2017normal, vogt2012silicene,yuhara2019graphene}. All these monolayers demonstrate a Dirac cone with a near to zero bandgap. On the other hand, group-V monolayers have been extensively studied in both theoretical and experimental works \cite{liu2014phosphorene,castellanos2014isolation,zhang2015atomically,ji2016two,pumera20172d}. Among them, phosphorene attracts huge research interest in 2D materials due to its proper bandgap, high carrier mobility, and excellent transport properties\cite{liu2014phosphorene}. After that, antimonene was introduced as an interesting 2D material with air stability\cite{ji2016two}.

The combinations of group-IV and V atoms can undertake superior electrical properties of both groups. Barreteau et al \cite{barreteau2016high} have built the bulk structure of layered SiP, SiAs, GeP, and GeAs. The layered configurations of these materials demonstrate that these materials can be exfoliated into 2D ­structures. The easy exfoliation of these materials have been approved experimentally \cite{kim2019thickness, jung2018two, yang2018highly, li20182d, cheng2018monolayered, jing2017gep3}. The monoclinic crystal of GeAs and SiAs has layered structure with C2/m space group \cite{cheng2018monolayered}. Monolayers of GeAs and SiAs can be exfoliated from the bulk counterparts due to low interlayer energy \cite{wu2016stability}. Both monolayers demonstrate a bandgap around 2 eV \cite{zhou2018geas}. The group IV-V monolayer compounds demonstrate a hexagonal lattice (V-IV-IV-V) with P6m2 space group \cite{ozdamar2018structural,lin2017single, ashton2016computational, shojaei2016electronic}. These hexagonal compounds are semiconductors expect CBi and PbN with metallic phase.  The structural stabilities and electronic properties of IV-V monolayers with A$_2$B$_2$ formula  (A=C, Si, Ge, Sn, Pb; B=N, P, As, Sb, Bi) have been analyzed theoretically  \cite{ashton2016computational, ozdamar2018structural}. Single-layer group IV-V compounds demonstrate fascinating photocatalytic activity \cite{shojaei2020silicon,zhang2020high,mortazavi2019two}, thermoelectric \cite{somaiya2020exploration, zhao2017geas2}, mechanical \cite{mortazavi2018anisotropic}, and electrical properties \cite{yang2018highly, li20182d, grillo2020observation}. The electrical properties of SiX (X=N, P, As, and Sb) monolayers demonstrate that these materials are semiconductors with an indirect band gap\cite{somaiya2020exploration}. These compounds have been reported as promising candidates for efficient thermoelectric applications.

\begin{figure*}
	\centering
	\includegraphics[width=1.0\linewidth]{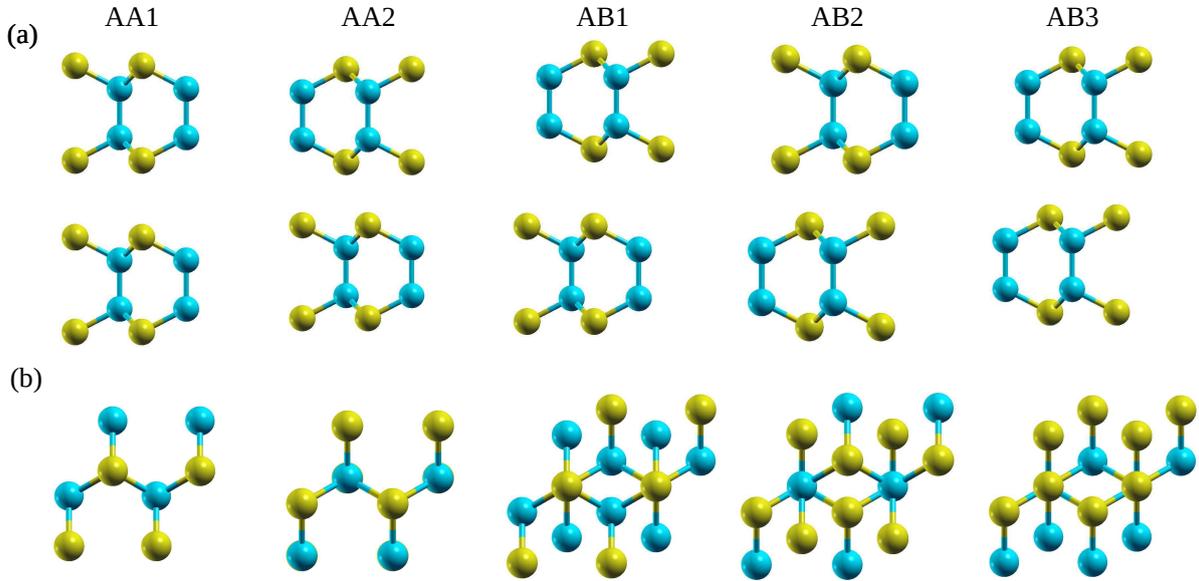}
	\caption{Bilayer SiX from (a) side, and (b) top view. The Si and X atoms are indicated by blue and yellow colors, respectively. }
	\label{fig:fig1}
\end{figure*}

Li et al \cite{li20182d} for the first time have exfoliated 2D GeP from the bulk monoclinic structure. Cheng et al \cite{cheng2018monolayered} have reported the exfoliation energy of SiP, SiAs, GeP and GeAs  which are of about 0.26 $\mathrm{J/m^2}$, 0.27 $\mathrm{J/m^2}$, 0.34 $\mathrm{J/m^2}$ and 0.37 $\mathrm{J/m^2}$, respectively. The exfoliation energy of SiP and SiAs are lower than graphite (0.32 $\mathrm{J/m^2}$), which confirms the experimental feasibility of their monolayers.

Field effect transistors (FET) based on IV-V have been introduced as a candidate for nano-electronic applications, however, their performance is limited by their low mobility. Guo et al \cite{guo2018few} have reported that the hole mobility of GeAs based FET at room temperature can reach 100 $\mathrm{cm^2/Vs}$. Monolayer GeP based FET is a p-type transistor and exhibits a I$\mathrm{_{on}}$/I$\mathrm{_{off}}$ ratio in the range of $\mathrm{10^4}$ \cite{li20182d}. The experimental results demonstrate that while the mobility of GeP can be enhanced when the thickness rises from ultrathin to bulk, the I$\mathrm{_{on}}$/I$\mathrm{_{off}}$ ratio is reduced\cite{kim2019thickness}.


Tuning the electrical and optical properties of the multilayer structures for their potential application in electro-mechanical devices, tunable photodetectors, and lasers is a challenge and can be done by changing stacking order, interlayer spacing, applying strain and electric field \cite{guo2008tuning, ghobadi2019normal, shamekhi2020band, touski2020interplay}. It has been shown that applying a vertical electric field can open a small bandgap even in bilayer graphene \cite{McCann2006bilayergraphene,castro2007biased}. Furthermore, it has been reported that a vertical electric field in the range of $\mathrm{0.2-0.3 V/\AA}$ leads to a semiconductor-to-metal transition in bilayer TMDs \cite{Ramas2011bilayertmd}. While this method is promising, it has practical problems such as the need for a very large electric field. On the other hand, it has been shown that the band structure of bilayer TMDs can be effectively modified by the application of vertical strain \cite{Bhattacharyya2012bilayertmd}

To the best of our knowledge, there is not a comprehensive study on the electrical and optical properties of different stacking orders of bilayer SiX (X=N, P, As, and Sb). The bilayer SiX will surely enrich the family of the 2D materials with fascinating electrical properties and So it is necessary to investigate the electrical and optical properties in these materials. In addition, applying out-of-plane strain is a powerful method to tune the bandgap and electrical properties of 2D materials. Therefore, in this work, the effect of vertical strain on the electrical properties of bilayer SiX (X=N, P, As, and Sb) is studied using density functional theory. Five different stackings are investigated for every compound and their electrical properties are discussed. The band
gaps of these materials decrease gradually with out-of-plane compressive strain and semiconductor-to-metal transition occurs at a specific pressure. This transition pressure depends on the stacking order of layers. This wide range (1.7–0.0 eV) bandgap tuning can be utilized in various applications.


\begin{table*}[t]
	\caption{The lattice constant ($a$), the interlayer distance ($d_{int}$), the distance between Si atoms ($d_{Si-Si}$), the distance between X atoms ($d_{X-X}$), Si-X ($d_{Si-X}$) bond length and elastic constants ($C_{11}$ and $C_{12}$) of SiX bilayers with different stacking orders.	\label{tab:tab1}}
	
	\begin{tabular}{p{1.5cm}p{1.5cm}p{1.5cm}p{1.5cm}p{1.5cm}p{1.5cm}p{1.5cm}p{1.5cm}p{1.5cm}p{1.5cm}}
		\hline
		\hline
		&Stacking order & $a$($\mathrm{\AA}$) & $d_{int}$($\mathrm{\AA}$) & $d_{Si-Si}$($\mathrm{\AA}$)  & $d_{X-X}$($\mathrm{\AA}$) & $d_{Si-X}$($\mathrm{\AA}$) & $E_b$(eV) & $C_{11}$(N/m) & $C_{12}$(N/m) \\
		\hline
		SiN & AA1 & 2.908 &	3.222 &	2.396 &	3.572 &	1.778 &	-1.341 & 558.45	& 136.12      \\
		& AA2 & 2.909 &	2.716 &	2.401 &	3.569 &	1.78 & -1.439  & 554.42 & 123.27     \\
		& AB1 & 2.907 &	3.188 &	2.398 &	3.574 &	1.779 &	-1.338  & 557.48 & 135.65       \\
		& AB2 & 2.907 &	2.932 &	2.403 &	3.583 &	1.78 & -1.383 & 519.66 & 123.91      \\
		& AB3 & 2.909 &	2.743 &	2.399 &	3.568 &	1.779 &	-1.416   & 555.37 &	138.46     \\
		\hline
		SiP & AA1 &3.536 &	3.596 &	2.357 &	4.416 &	2.287 &	-1.51 & 288.98 & 57.09  \\
		& AA2 & 3.537 &	3.039 &	2.357 &	4.415 &	2.287 &	-1.609 & 287.86 & 56.79  \\
		& AB1 & 3.537 &	3.587 &	2.356 &	4.412 &	2.287 &	-1.509 & 288.72 & 56.01   \\
		& AB2 & 3.541 & 2.998 &	2.361 &	4.417 &	2.289 &	-1.639 & 279.62 & 55.97     \\
		& AB3 & 3.54 & 3.002 &	2.359 &	4.416 &	2.289 &	-1.629 & 288.36 & 53.94 \\
		\hline
		SiAs & AA1 & 3.683 & 3.71 &	2.348 &	4.576 &	2.402 &	-1.809 & 254.50 & 54.86     \\
		& AA2 & 3.686 &	3.106 &	2.346 &	4.576 &	2.403 &	-1.936 & 252.78 & 51.62 \\
		& AB1 & 3.683 &	3.713 &	2.347 &	4.576 &	2.402 &	-1.809 & 254.34 & 54.42 \\
		& AB2 &  3.693 & 2.994 & 2.347 & 4.565 & 2.405 & -1.965 & 241.08 & 54.63 \\
		& AB3 &  3.688 & 3.065 & 2.347 & 4.574 & 2.404 & -1.953 & 252.88 & 48.73   \\
		\hline
		SiSb & AA1 &  3.978	 & 4.074 & 2.342 &	4.819 &	2.609 &	-2.042 & 197.94 & 49     \\
		& AA2 &  3.982 & 3.273 & 2.341 & 4.814 & 2.611 &-2.228  &194.86 & 41.31     \\
		& AB1 &  3.978 & 4.069 & 2.342 & 4.819 & 2.609 & -2.043 & 197.63 & 49.19   \\
		& AB2 &  3.994 & 3.145 & 2.342 & 4.799 & 2.614 & -2.277 &190.01	& 32.75  \\
		& AB3 &  3.987 & 3.222 & 2.34 &	4.81 &	2.612 &	-2.252  & 196.06 & 34.95 \\
		\hline
		\hline
	\end{tabular}
\end{table*}

\begin{figure*}
	\centering
	\includegraphics[width=1.0\linewidth]{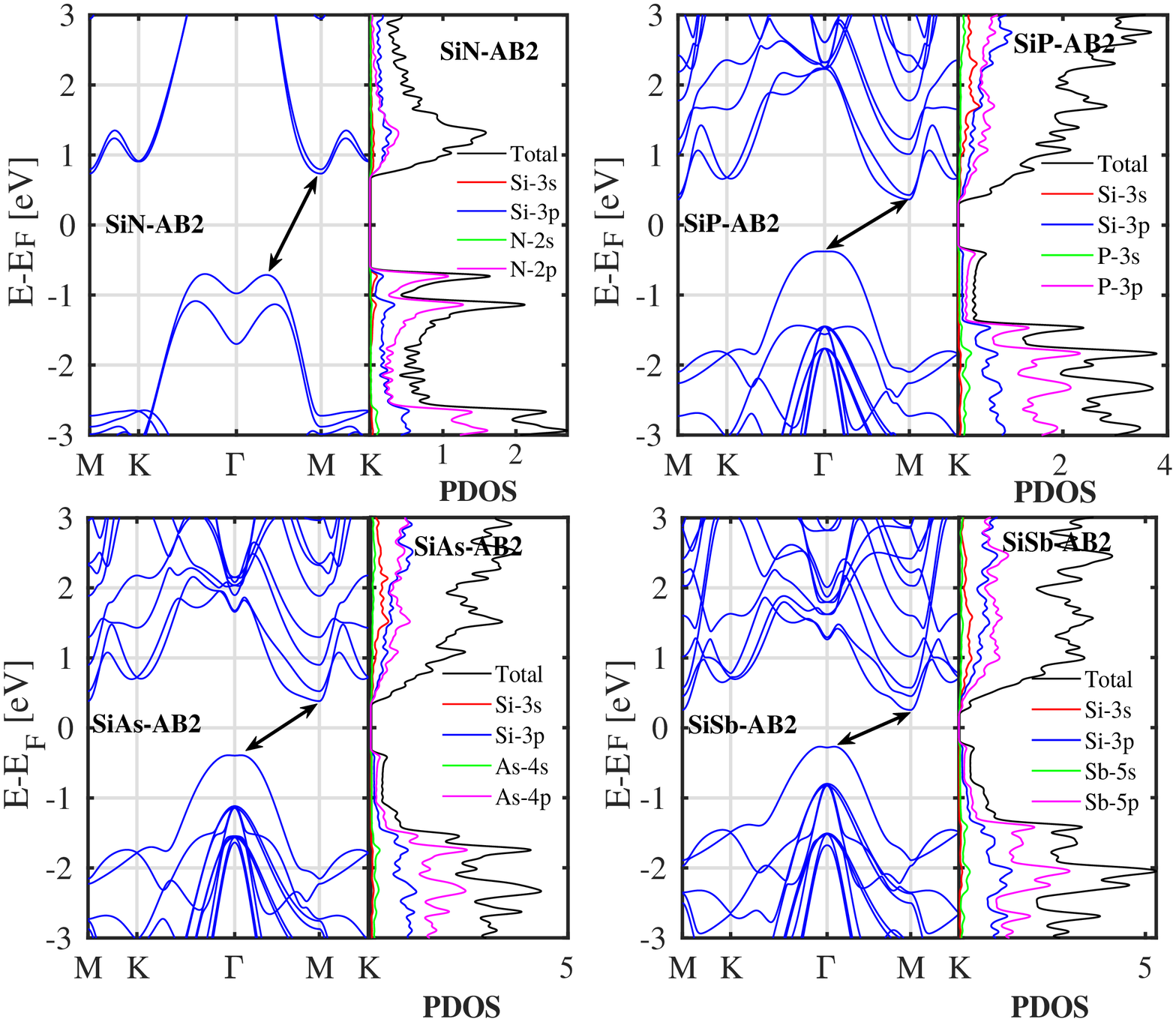}
	\caption{The band structures and PDOS of bilayer SiX with AB2 stacking order. }
	\label{fig:fig2}
\end{figure*}

\section{Computational details}
Density functional calculations are performed using the SIESTA package \cite{soler2002siesta}. The generalized gradient approximation (GGA) with the Perdew-Burke-Ernzerhof (PBE) \cite{perdew1981self} functional is utilized for the exchange-correlation term. A Monkhorst-Pack k-point grid of $21\times21\times1$ is chosen for the unit-cell. The energy cutoff is set to be 200 Ry and a double-$\zeta$ plus polarization basis-set is used. The total energy is converged to better than $10^{-5}$ eV. The geometries are fully relaxed until the force on each atom is less than 0.01 eV/$\mathrm{\AA}$. A vacuum region of 30 $\mathrm{\AA}$ is added to avoid interactions in the normal direction. The van der Waals (vdW) interaction between layers is treated using Grimme's correction to the PBE functional \cite{grimme2006}. To visualize the atomic structures, XCrySDen package has been used \cite{kokalj2003computer}. The vertical strain is defined as, 

\begin{equation}
\varepsilon=\frac{(d-d_0)}{d_0} 
\end{equation}

where $d_0$ and $d$ are the equilibrium and deformed interlayer distances, respectively. The applied pressure (P) is calculated from the energy cost per unit area for decreasing the interlayer distance by following equation \cite{Bhattacharyya2012bilayertmd},

\begin{equation}
P=\frac{(E-E_0)}{(d_0-d)A} 
\end{equation}

where $A$ is the area of the unit cell, and $E_0$ and $E$ are the energies of the equilibrium and deformed structures. The effective mass of the carriers is calculated by using the following equation \cite{ghobadi2020electrical,touski2020electrical},
\begin{equation}
m^*=\hbar^2/\left(\partial^2E/\partial k^2\right)
\end{equation}
Here, $\hbar$ is the reduced Planck constant, E and k are the energy and wave vector of conduction band minimum and valence band maximum. The absorption coefficient is calculated using the energy dependent dielectric functions. In the optical absorption simulation, the absorption coefficient can be calculated by \cite{huang2019optical}
\begin{equation}
\alpha(\omega)=\sqrt{2}\frac{\omega}{c}[\sqrt{\varepsilon_1^2(\omega)+\varepsilon_2^2(\omega)}-\varepsilon_1(\omega)]^{\frac{1}{2}}
\end{equation}

where $\omega$, $c$, $\varepsilon_1(\omega)$, and $\varepsilon_2(\omega)$ are the angular frequency of light, the speed of light, and the real and imaginary parts of the dielectric
function, respectively.

\begin{table*}[t]
	\caption{The electrical properties of the different stacking orders of bilayer SiX. The band gap ($E_g$) is in eV unit. The strain ($\varepsilon_{tran}$) and pressure ($P_{tran}$) required for semiconductor-to-metal transition are in percent and GPa, respectively. The effective masses at $\Gamma$-point of valence band, and K- and M-valleys of the conduction band are in m$_0$ unit. The Mexican-hat energy ($E_{M\Gamma}$) and coefficient ($M_{\Gamma}$) at $\Gamma$-point of the valence band are in eV and $\mathrm{eV\AA}$ unit, respectively.	 \label{tab:tab2}}
	
	\begin{tabular}{p{1.1cm}p{1.5cm}p{1.1cm}p{1.1cm}p{1.1cm}p{1.1cm}p{1.1cm}p{1.1cm}p{1.1cm}p{1.1cm}p{1.1cm}p{1.1cm}p{1.1cm}}
		\hline
		\hline

		&Stacking order & $E_g$ & $\varepsilon_{tran}$ & $P_{tran}$& $m^{c,*}_{K\rightarrow M}$ & $m^{c,*}_{K\rightarrow \Gamma}$  & $m^{c,*}_{M\rightarrow \Gamma}$ & $m^{c,*}_{M\rightarrow K}$ &  $m^{v,*}_{\Gamma\rightarrow K}$ & $m^{v,*}_{\Gamma\rightarrow M}$ &$E_{M\Gamma}$ & $M_{\Gamma}$ \\
		\hline
		SiN & AA1 &1.401 &-18&16.14 &0.585&	0.553&		1.081&	0.383	 &1.091&	0.989&0.331 &0.696	      \\
		& AA2 &1.232 &-36 &42.47 &0.538&	0.535&		1.045&	0.371	 &1.143&	1.051&0.206 &0.506	    \\
		& AB1 & 1.413 &-20 &19.73 &0.588&	0.56&		1.081&	0.377	 &1.099&	0.988&0.324 &0.681	       \\
		& AB2 & 1.434&-20 &16.91 &0.669&	0.57&		1.264&	0.416	 &1.163&	1.079&0.264 &0.596   \\
		& AB3 &  1.262& -14 &10.82 &0.634&	0.589&		1.09&	0.388	 &1.19&	1.087&0.221 &0.527	   \\
		\hline
		SiP & AA1&1.064 &-9 &4.04 &0.434&	0.461 &  	2.321&	0.141&	1.729&	1.402&0.021 &0.134	  \\
		& AA2&0.768 &-6 &2.50 &	0.382&	0.423 &	2.839&	0.142&		1.436&	1.289 &0.002 &0.025	   \\
		& AB1 &1.05 &-8 & 3.42&	0.434&	0.467 &  	2.144&	0.14&		1.683&	1.364&0.021 &0.138  \\
		& AB2 &0.742 &-6 &1.14 &0.501&	0.496 & 2.412&	0.134&		1.577&	1.318&0.001 &0.017	     \\
		& AB3 &0.731 &-6 &1.96 &	0.484&	0.52  &	2.338&	0.138&		1.459&	1.282&0.001 &0.019	 \\
		\hline
		SiAs & AA1  &1.258& -10& 4.47&0.428	&0.46&	6.237&	0.133&		1.848&	1.461&0.009 & 0.096  \\
		& AA2 & 0.92& -9&4.1 &0.379&	0.427&		2.804&	0.136&		1.004&	0.884&0.008 &0.101	 \\
		& AB1 &1.263 &-10 &4.36 &0.427&	0.463&		5.468&	0.134&		1.829&	1.451&0.01 &0.097	 \\
		& AB2 &0.773 &-6 &1.65 &0.503&	0.518&		2.953&	0.125&	0.978&	0.822&0.005 &0.068 \\
		& AB3 & 0.879 &-8 &2.87 &0.47&	0.506&		5.621&	0.132&	0.999&	0.86&0.007 &0.089   \\
		\hline
		SiSb & AA1 & 0.995 &-10 &3.99 &0.411	&0.446		&0.411	&0.2	&0.445&0.459&0.0 &0.0     \\
		& AA2 & 0.476&-4 &1.36 &0.351	&0.408		&0.38	&0.176	&0.782	&0.685&0.019 &0.225	      \\
		& AB1 & 1.086& -11&4.54 &0.419	&0.451	&0.443	&0.243		&0.444	&0.459&0.0 &0.0	   \\
		& AB2 & 0.524 &-4 &0.92 &0.392	&0.653		&0.875	&0.131		&0.707	&0.59&0.012 &0.155	  \\
		& AB3 & 0.568 & -5&1.51 &0.415	&0.445	&0.626	&0.166		&0.748	&0.644& 0.016&0.188	 \\
		\hline
		\hline
	\end{tabular}
\end{table*}

\section{Results and discussion}
The schematic of five different stackings has been displayed in top and side views in Fig. \ref{fig:fig1}. Two stacking categories are AA and AB which in AA stackings, the top layer is located exactly on the underlying layer, and in AB stackings, the top layer is shifted relative to the bottom layer. The structural properties of the bilayers are listed in Table \ref{tab:tab1}. The lattice constants are almost equal for the five stackings of each material which implies that stacking order has a negligible effect on the lattice constant. On the other hand, interlayer distance highly depends on the stacking configuration. The AA2 stacking of SiN has the lowest interlayer distance, whereas, the lowest one is for AB2 stacking in other compounds. The highest $d_{int}$ is for AA1 and AB1 stackings and the interlayer distance is approximately the same in these two stackings. $d_{Si-Si}$, $d_{Si-X}$ and $d_{X-X}$ similar to lattice constant demonstrate a low dependency on the stacking configuration. Binding energy can be computed as\cite{ghobadi2019normal}, 
\begin{equation}
E_b=E_{Bilayer}-2\times E_{Monolayer}
\end{equation}
where $E_{Bilayer}$ and $E_{Monolayer}$ are the total energy of bilayer and monolayer SiX, respectively. The binding energy decreases with the atomic number of group-V elements and the heavier compounds display lower binding energy. The lowest binding energy which corresponds to the most stable structure is for AB2 configuration in all compounds except SiN that AA2 has the lowest binding energy. As one can observe, the binding energy highly depends on the interlayer distance. The lower interlayer distance results in the lowest binding energy. The elastic constants, $C_{11}$, $C_{22}$ and $C_{12}$ are also studied. $C_{22}$ is the same as $C_{11}$ and have not been written in the table. The stability of these configurations is approved by the born stability criteria as: 
$0<C_{11}$, $0<C_{22}$ and $C_{12}<C_{11},C_{22}$ \cite{Mouhat2014stability}. The values of $C_{11}$ and $C_{12}$ decrease with increasing the atomic number. Bilayer SiN demonstrates the highest elastic constants that are about 50$\%$ and 300$\%$ larger than bilayer graphene and MoS$_2$, respectively \cite{Guoxin2014Graphene, Duerloo2012mos2}. $C_{11}$ shows a dependency with interlayer distance and binding energy in most cases and the highest $C_{11}$ belongs to the stacking with the highest interlayer distance and largest binding energy.

\begin{figure}
	\centering
	\includegraphics[width=0.9\linewidth]{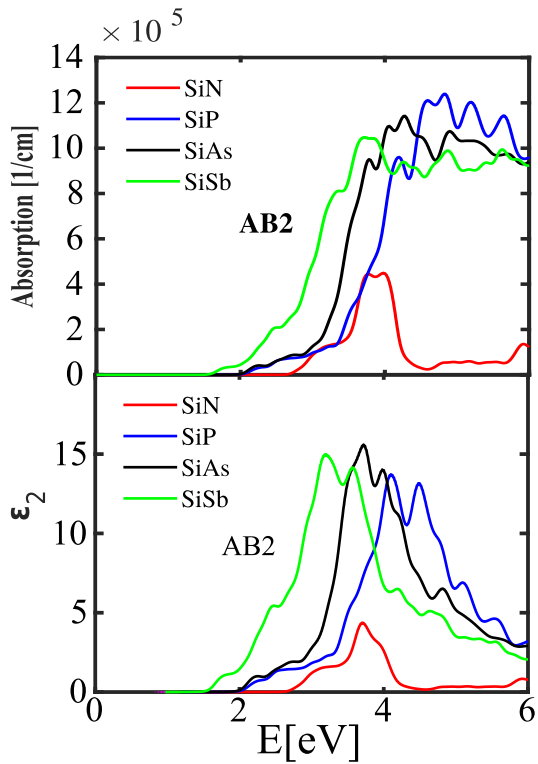}
	\caption{Absorption coefficient and imaginary part of the dielectric function of AB2 stacked bilayer SiX. }
	\label{fig:fig3}
\end{figure}

\begin{figure*}
	\centering
	\includegraphics[width=1.0\linewidth]{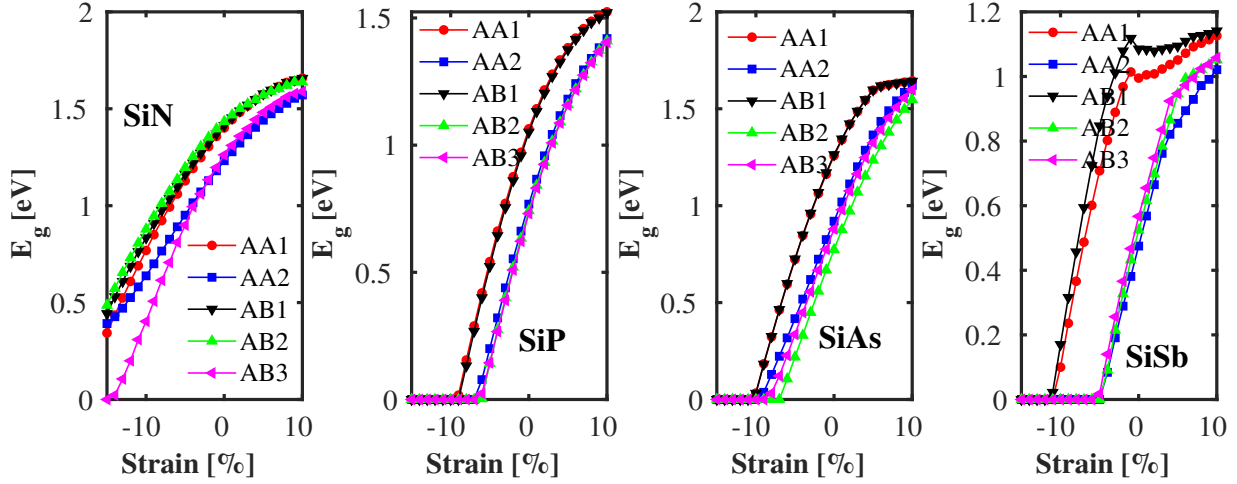}
	\caption{The band gap variation of different stacking orders of bilayer SiX as a function of vertical strain. }
	\label{fig:fig4}
\end{figure*}

\begin{figure}
	\centering
	\includegraphics[width=1\linewidth]{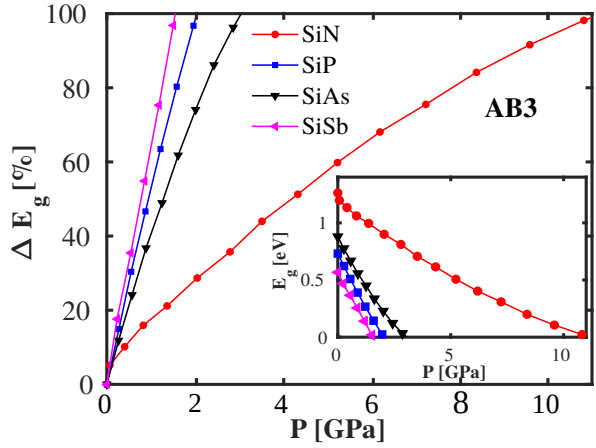}
	\caption{Percentage change in bandgap with the applied pressure for AB3 stacked bilayer SiX as a function of the applied pressure. The inset figure shows the value of the bandgap versus the applied pressure.}
	\label{fig:fig5}
\end{figure}

\begin{figure*}
	\centering
	\includegraphics[width=1.0\linewidth]{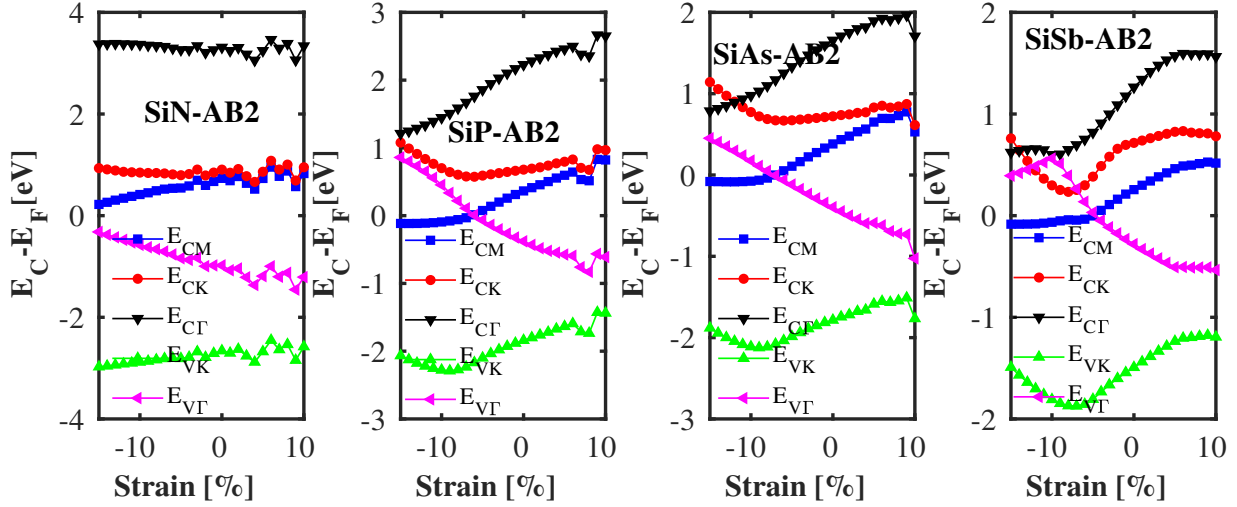}
	\caption{The energy of the valleys in the conduction and valence bands for various strain in AB2 stacking of bilayer SiX.}
	\label{fig:fig6}
\end{figure*}

\begin{figure*}
	\centering
	\includegraphics[width=1.0\linewidth]{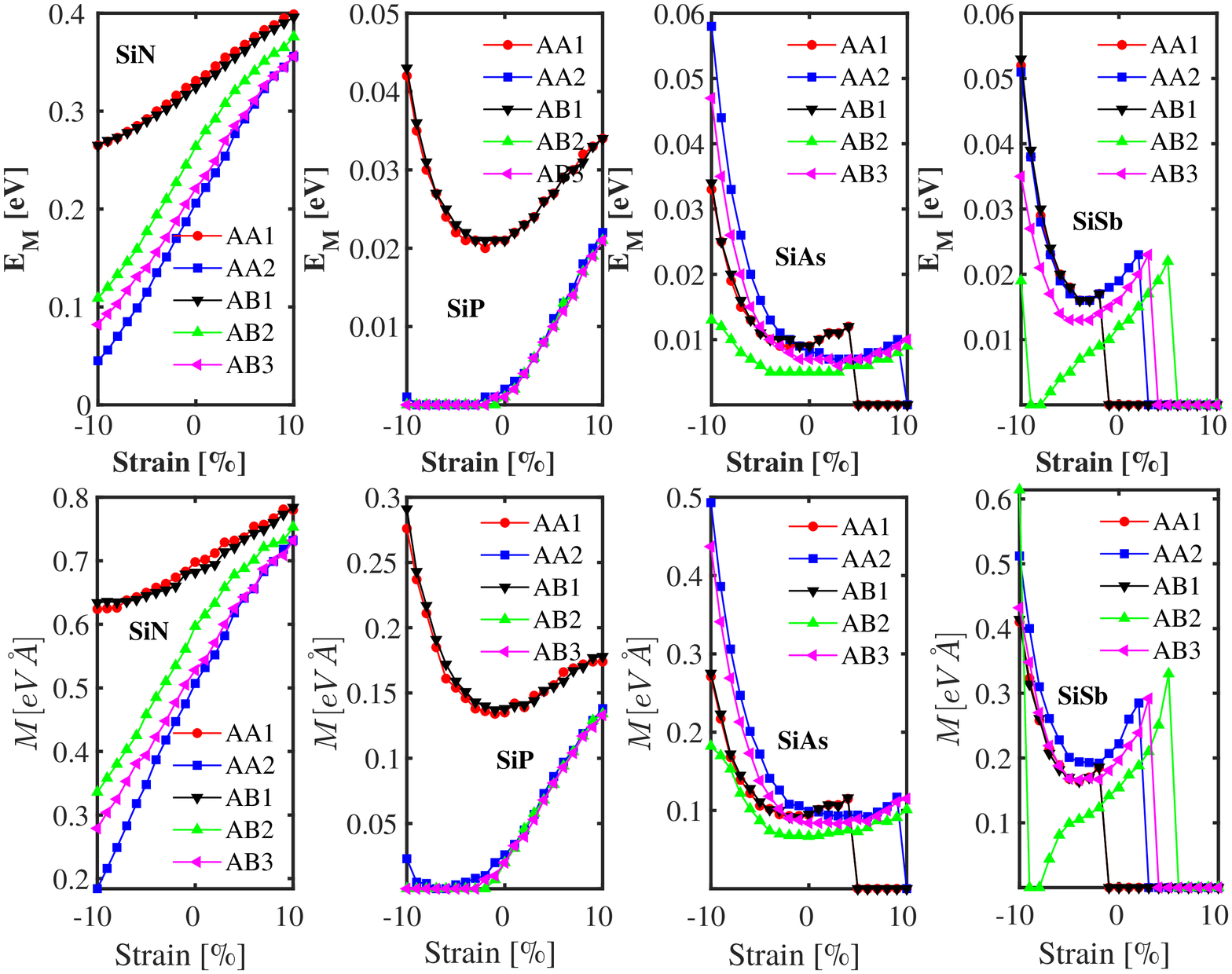}
	\caption{The Mexican-hat energy and coefficient of SiX bilayers as a function of out-of-plane strain. The top and bottom rows stand for energy and coefficient of the Mexican-hat, respectively.}
	\label{fig:fig7}
\end{figure*}

The band structures of the bilayer SiX with AB2 configuration are depicted in Fig. \ref{fig:fig2}. The AB2 stacking which is the most stable configuration is selected as a sample. All of them are indirect semiconductor where conduction band minimum (CBM) is located at M-valley and valence band maximum (VBM) happens at $\Gamma$ or a point near to $\Gamma$ (we have called this point as $\Gamma^*$). The creation of Mexican-hat is the reason of the moving of VBM from $\Gamma$- to  $\Gamma^*$-point. The values of the band gaps for different stackings of SiX compounds are listed in Table \ref{tab:tab2}. The size of the band gaps of the bilayer SiN are distributed from 1.232eV in AA2 to 1.434eV in AB2 stacking. The band gap of Bilayer SiN displays a low dependency on the stacking order. On the other hand, the band gap of bilayer SiSb highly depends on the stacking orders. The band gap of AB1 stacking of SiSb is 1.086 eV that is approximately two times as large as the band gap of the AA2 stacking. The band gap of heavier compounds have a higher dependency on the stacking order. We also observed a high dependency of the band gap on the stacking order in bilayer antimonene \cite{touski2020interplay}.  
The highest band gap in Bilayer SiN is 1.434eV for AB2 stacking. After AB2 stacking, AB1 and AA1 stackings have the highest band gap. In three other materials, SiP, SiAs and SiSb, AA1 and AB1 stackings approximately have the same bandgap and demonstrate the highest band gap. The lowest band gap is one of the AA2, AB2 or AB3 stackings that is different for various compounds. For example, the lowest band gap in SiN is 1.232 eV in AA2 stacking, whereas, in SiAs is for AB2 stacking. 

All structures exhibit the Mexican-hat dispersion in the top of the valence band that is more noticeable in bilayer SiN. The value of the Mexican-hat coefficient can be obtained with \cite{ariapour2019spin}: $M=\Delta E/\Delta K$, where $\Delta E$ and $\Delta K$ are the energy and momentum difference between the $\Gamma$-point and the valence band maximum. The values of the Mexican-hat energies and coefficients for different stackings of SiX compounds are listed in Table \ref{tab:tab2}. Bilayer SiN demonstrates the highest Mexican-hat coefficient of 0.696 eV$\mathrm{\AA}$ at AA1 stacking. The different stackings display different Mexican-hat coefficients. Mexican-hat energies for bilayer SiN are distributed from 0.206 to 0.324eV. On the other hand, despite the negligible $E_M$ of other compounds, some stackings demonstrate a relatively high Mexican-hat coefficient.

For better understanding of the contribution of the atoms and orbitals on the band structure, partial density of states (PDOS) for AB2 stackings are plotted along with the band structures in Fig. \ref{fig:fig2}. The singularity in the valence band especially in SiN is obvious. There also exist the singularity in the valence band of the other compounds but their values are small. These singularities are come from the Mexican-hat in the VBM. As one can observe, the p orbitals of two atoms have the main contribution to the both conduction and valence band edges.  

While the conduction band minimum is located at M-valley, energy of the K-valley in some of the configurations is close to M-valley and K-valley contributes to the conduction band minimum. On the contrary, the valence band maximum is located at $\Gamma^*$-point  and the energy of the M- and K-points are much lower. The effective masses for the M- and K-valleys of the conduction band and $\Gamma^*$-point of the valence band are listed in Table \ref{tab:tab2}. Two effective masses at K-valley in the conduction band ($m^c_{K\rightarrow \Gamma}$ and $m^c_{K\rightarrow M}$) demonstrate almost the same values in the most of configurations. On the other hand, M-valley presents two different effective masses, $m^c_{M\rightarrow \Gamma}$ and $m^c_{M\rightarrow K}$. $m^c_{M\rightarrow \Gamma}$ is approximately three times as large as $m^c_{M\rightarrow K}$ in bilayer SiN. Their ratio reaches to more than one order of magnitude in SiP, and SiAs shows the highest difference between these two effective masses. On the other hand, SiSb behaves differently and the difference between these two effective masses highly depends on the stacking order. Two effective masses are also calculated at $\Gamma$-point of the valence band that are approximately the same. These materials exhibit a high hole effective mass. SiP displays the highest hole effective mass, whereas, SiSb has the lowest one close to its electron effective mass.

The imaginary part of the dielectric functions ($\varepsilon_{2}$) determines the optical absorption \cite{madelung2012introduction,soleimani2020effects}. For this reason, $\varepsilon_{2}$ as a function of energy for four compounds are depicted in Fig. \ref{fig:fig3}. $\varepsilon_{2}$ is zero for energy lower than 2 eV and is enhanced after that. SiSb with lower bandgap demonstrates $\varepsilon_{2}$ at lower energy whereas in SiN with the highest bandgap, $\varepsilon_{2}$ starts at higher energy. As one can observe, the band gaps of these compounds are distributed in the range of 0.476 to 1.434eV that are lower than those energies that $\varepsilon_{2}$ rises. The peaks in the conduction and valence bands of DOS determines optical absorption. In SiN, $\varepsilon_{2}$ is compatible with absorption coefficient and is associated with the peaks of DOS in the valence band and the conduction band. Three other compounds, SiP, SiAs and SiSb, have a single band in the valence band edge that limits the band gap. As one can see, this band doesn't create a high DOS and doesn't highly contribute to the optical absorption. DOS displays a maximum after second band and these bands contribute to $\varepsilon_{2}$. Bilayer SiP, SiAs and SiSb demonstrate the same maximum value of $\varepsilon_{2}$ with a shift in energy, whereas, SiN shows much lower value. The optical absorption as a function of energy for all bilayers are shown in Fig. \ref{fig:fig3}. The optical absorption is compatible with $\varepsilon_{2}$.

Out-of-plane strain has been proposed as a powerful method to modify the electrical properties of bilayers or hetero-structures \cite{Bhattacharyya2012bilayertmd, ghobadi2019normal, shamekhi2020band}. We have applied the out-of-plane strain to all stackings and their electrical properties are studied. The variation of the band gap of the compounds as a function of vertical strain is plotted in Fig. \ref{fig:fig4}. The band gaps are enhanced with tensile strain. On the other hand, vertical compressive strain decreases the band gap values and semiconductor-to-metal transition occurs at a specific strain in compressive regime. The required strain for phase transition ($\varepsilon_{trans}$) and its counterpart pressure ($P_{trans}$) are listed in Table \ref{tab:tab2}. $\varepsilon_{trans}$ is too high for bilayer SiN,for example, it reaches to -36$\%$ for AA2 stacking that needs 42.47 GPa of pressure. $\varepsilon_{trans}$ depends on the value of the band gap and a larger band gap results in a higher $\varepsilon_{trans}$. Only AA2 stacking of SiN doesn't obey from this theorem. $P_{trans}$ of bilayer SiN are distributed between 10 and 20 GPa, except AA2 stacking. SiP, SiAs and SiSb demonstrate a lower $\varepsilon_{trans}$ and $P_{trans}$. One can observe, the band gap of all stackings of bilayer SiN are close to each other under different stains and just AB3 stacking exhibits a lower band gap at compressive strain. In SiP, SiAs and SiSb, the band gap of AA1 and AB1 configurations vary close to each other. Furthermore, the three structure with more stability (AA2, AB2 and AB3) vary similarly with strain. The difference between these two groups increases for heavier compound. As one can observe, SiSb demonstrates a large difference between these two groups.

These structures can be introduced as a pressure sensor. In order to investigate the feasibility of semiconductor to metal transition and possible application of the structures as a pressure sensor in experiments, the applied pressure (P) is calculated and the sensitivity of the band gap on the applied pressure is studied. The band gap variation versus the applied pressure is plotted in Fig. \ref{fig:fig5} for AB3 stacking. The plot is almost linear for all of the structures and the pressure range is easily achievable experimentally. SiSb displays the highest sensitivity to the pressure and the bandgap closes at a low pressure. The band gap of SiSb closes at a pressure lower than 2GPa. The sensitivity of the band gap decreases with decrement of X atomic number. SiN demonstrates the lowest band gap sensitivity but remains a semiconductor until the pressure of 10GPa. The transition pressure decreases as the atomic number of X atom increases. This is due to the increased delocalization of the atomic orbitals, which leads to reduced interaction between Si and X atoms and a lower transition pressure. This trend is also observable in the band gap variation with X atomic number [See Table \ref{tab:tab2}].

The energy of the effective valleys in the conduction and valence bands for various strains are plotted in Fig. \ref{fig:fig6}. The behavior of all stackings are the same and AB2 stacking is plotted as a sample. The figures demonstrate the CBM and VBM are located at M- and $\Gamma$-points, respectively. The energies of K-valleys in the conduction band ($E_{CK}$) become closer to $E_{CM}$ under tensile strain. However, the energy of $\Gamma$-valley in the conduction band and K-point in the valence band are far from the CBM and VBM, respectively. Therefore, these two valleys don't contribute to electrical properties. $E_{CM}$ decreases with applying compressive strain and in the same time, $E_{V\Gamma}$ rises. These two bands intersect at $\varepsilon_{trans}$. Therefore, the CBM and VBM get closer to Fermi level and electron and hole density increase exponentially that results in increment of the current.

The compounds demonstrate Mexican-hat dispersion at the valence band that can be affected by the strain. Mexican-hat energy and coefficient as a function of strain are plotted in Fig. \ref{fig:fig7}. SiSb, SiAs and SiSb display a comparable Mexican-hat coefficient with SiN, whereas, their Mexican-hat energies are much lower than SiN. One can observe that the value of the Mexican-hat energy in SiN is approximately one order of magnitude larger than the others that is compatible with Table \ref{tab:tab2}. Mexican-hat energy and coefficient behave similarly in a compound. For example, in the bilayer SiN Mexican-hat energy and coefficient increase with the tensile strain and decrease in compressive regime. However, Mexican-hat properties in other compounds show a minimum at equilibrium and increase with applying both compressive and tensile strains. Mexican-hat vanishes at tensile strain for SiAs and SiSb bilayers. AA1 and AB1 stackings demonstrate a larger Mexican-hat properties than the other stackings in the SiN and SiP and their differences rises at compressive stain.

\section{Conclusion}  
The structural, electrical and optical properties of five different stacking orders of bilayer SiX (X=N, P, As and Sb) are studied. All these materials are indirect semiconductors where the CBM and VBM are located at M- and $\Gamma^*$-points, respectively. The Mexican-hat is obvious from the band structures and the singularity in the DOS confirms the existence of the Mexican-hat. SiN stackings demonstrate a considerable Mexican-hat dispersion, whereas, it is negligible for other compounds. In the following, out-of-plane strain has been applied to tune the electrical properties. The band gap increases with a rise in the tensile strain and a semiconductor-to-metal transition occurs at compressive strains. SiSb demonstrates the highest band gap sensitivity to the pressure, whereas, SiN has the lowest band gap sensitivity and closes at higher level of pressure. At high tensile strain, the energy of the K-valley gets closer to the M-valley and contributes to the CBM. SiN displays a high Mexican-hat coefficient that increases with tensile strain and decreases with compressive strain. Other compounds also indicate a high Mexican-hat coefficient but the energy of the Mexican-hat is small for them. The optical absorption is also studied where the peaks in the conduction and valence bands of DOS determines optical absorption. The single band in the valence band edge of SiP, SiAs and SiSb has a low DOS and a little contribution to the optical absorption.

\end{document}